\documentclass[aps,prl,twocolumn,groupedaddress,superscriptaddress,amsmath,amssymb,
]{revtex4-2}

\usepackage{graphicx, float}

\begin{document}

\title{Writhing and hockling instabilities in twisted elastic fibers
}

\author{Adam Fortais}
\affiliation{Department of Physics and Astronomy, McMaster University, 1280 Main Street West, Hamilton, Ontario, L8S 4M1, Canada} 
\author{Elsie Loukiantchenko}
\affiliation{Department of Physics and Astronomy, McMaster University, 1280 Main Street West, Hamilton, Ontario, L8S 4M1, Canada} 
\author{Kari Dalnoki-Veress}
\email[]{dalnoki@mcmaster.ca}
\affiliation{Department of Physics and Astronomy, McMaster University, 1280 Main Street West, Hamilton, Ontario, L8S 4M1, Canada} 
\affiliation{UMR CNRS Gulliver 7083, ESPCI Paris, PSL Research University, 75005 Paris, France.
}

\date{\today}

\begin{abstract}
The buckling and twisting of slender, elastic fibers is a deep and well-studied field. A slender elastic rod that is twisted with respect to a fixed end will spontaneously form a loop, or \emph{hockle}, to relieve the torsional stress that builds. Further twisting results in the formation of plectonemes -- a helical excursion in the fiber that extends with additional twisting. Here we use an idealized, micron-scale experiment to investigate the energy stored, and subsequently released, by hockles and plectonemes as they are pulled apart, in analogy with force spectroscopy studies of DNA and protein folding. Hysteresis loops in the snapping and unsnapping inform the stored energy in the twisted fiber structures.
\end{abstract}

\maketitle


\section{\label{sec:intro}Introduction}
Take a cord, twist one end with respect to the other, then relax the tension in the cord. We have all likely encountered the spontaneous looping and subsequent helix that will form with additional twisting. Hockling, or the buckling and looping of a twisted rod, is a well known and intensely studied phenomenon
~\cite{yu_bifurcations_2018,champneys_multiplicity_1996,thompson_helix_1996,coyne_analysis_1990,dias_wunderlich_2015,van_der_heijden_instability_2003,van_der_heijden_lock-tape-like_1999,clauvelin_matched_2009,gazzola_forward_2018,goss_experiments_2005,stump_hockling_2000,ross_cable_1977,mahasol2005,nizette_towards_1999,neukirch_classification_2002,PhysRevLett.115.118302}. When a cord is twisted, a significant amount of elastic energy can be stored in the twists of the fiber. By introducing slack in the fiber, this stored energy can be released by untwisting, however, if the rotation at the ends of the cord are fixed, untwisting is accompanied by bending into a hockle or plectoneme (a double-helix structure terminating in a loop) since the ends are fixed. When this occurs, hockling is preceded by a modified Euler buckling that results in sinusoidal buckles, eventually coarsening into a single loop~\cite{champneys_multiplicity_1996,thompson_helix_1996,coyne_analysis_1990}. 

Because hockling can result in damage to cables, determining the criteria for hockling is of practical engineering concern. Research has focused on experiments using braided cables, nickel-titanium (nitinol) rods, and plastic fibers with thicknesses ranging from millimeters to centimeters~\cite{wada_structural_2016,fraser_theory_2008,ermolaeva_hockling_2008,habibi_coiling_2007,ross_cable_1977,mahasol2005,thompson_helix_1996,coyne_analysis_1990,van_der_heijden_instability_2003,goss_experiments_2005}. However, performing these experiments at such large scales leads to complicating factors such as gravitational sagging, material defects, and non-uniformity. Furthermore, few experimental studies have explored the removal of hockles~\cite{goss_experiments_2005,yabuta_submarine_1984,ross_cable_1977,kojima_cable_1982,ermolaeva_hockling_2008}. Additionally, the choice of material in these studies are often prone to plastic deformation at low strains, limiting the study of hockle removal to small degrees of twist. 

Beyond hockling -- the creation of the first loop -- a highly twisted rod may begin writhing, where the associated rotation of the loop results in the formation of a plectoneme~\cite{thompson_helix_1996,champneys_multiplicity_1996,van_der_heijden_instability_2003,gazzola_forward_2018,mahasol2005,fraser_theory_2008,marko_competition_2012,fraser_equilibrium_1998,purohit_plectoneme_2008,clauvelin_elasticity_2009,neukirch_analytical_2011,kulic_equation_2007,marko_global_2013,smith_predicting_2008,daniels_discontinuities_2009,chamekh_stability_2014,brutzer_energetics_2010}. While this is less common in engineering applications, it occurs frequently in biological systems like DNA and plant tendrils~\cite{silverberg_3d_2012,gerbode_how_2012,tanaka_elastic_1985,coleman_theory_1995,hoffman_link_2003,starostin_three-dimensional_1996,gromiha_anisotropic_1996,cherstvy_looping_2011,dobrovolskaia_twist_2010,marko_competition_2012}. However, it is difficult to study this process in-vitro, and in the particular case of DNA, thermal fluctuations may both initiate plectoneme formation as well as introduce noise into any potential force measurements~\cite{goyal_nonlinear_2005,purohit_plectoneme_2008,lipfert_magnetic_2010,mosconi_measurement_2009,clauvelin_elasticity_2009,kulic_equation_2007,marko_statistical_1995,stump_writhing_1998,ganji_intercalation-based_2016}. Gaining a deeper understanding of twisted fibers could help in developing bio-inspired smart materials~\cite{studart_bioinspired_2014,hu_buckling-induced_2015,gazzola_forward_2018}.

In this study, uniform, cylindrical, elastic fibers with diameters on the order of $\sim10 \ \mu$m are used to experimentally investigate the hockling and writhing phenomena. Much like force spectroscopy measurements carried out with DNA  and magnetic tweezers~\cite{lipfert_magnetic_2010,mosconi_measurement_2009}, here on  larger length scales we employ a micro-pipette deflection technique~\cite{Colbert2009,Backholm2019MicropipetteFS} to quantify the tension, twisting, and bending energies in the system as a fiber hockles, writhes, and is pulled apart again. Extending the work of Ross and Yabuta, we first derive exact, material-independent hockling and hockle-removal criteria~\cite{ross_cable_1977,yabuta_submarine_1984}. The expression derived is purely geometric with no fit parameters, and accurately describes experiments performed with fibers of various sizes. We then focus on the formation and removal of plectonemes from a twisted fiber, with the latter revealing an especially rich force response. 
\section{\label{sec:Experiment}Experiment}
\begin{figure}[]
\centering
\includegraphics[width=0.9\columnwidth]{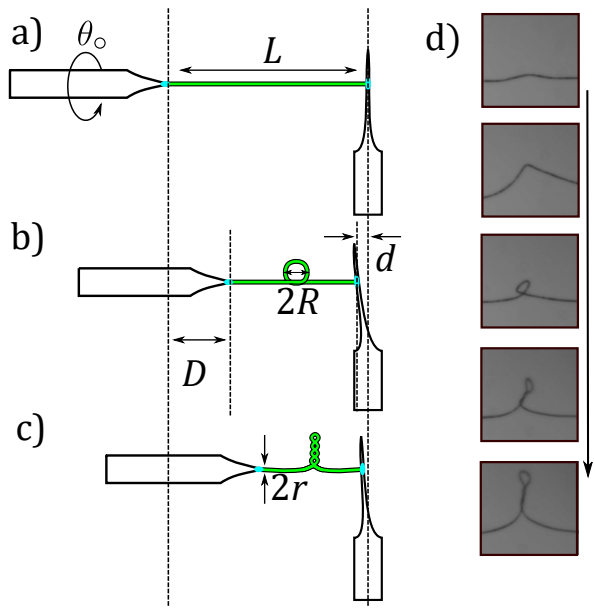}
\caption{\label{schem}
a) A fiber with radius $r$ and length $L$ was glued to two posts, a mobile post (left) and a force-transducer post (right). The left post is rotated by an angle $\theta_{\circ}$,  and moved toward the right post. b) As the posts move together, the fiber untwists with the formation of a loop. c) Upon further decreasing the distance a plectoneme forms. The process is then reversed and the glass pipettes are separated. Tension in the fiber is measured by observing the deflection $d$ of the force-transducer pipette and $D>>d$. d) Optical image sequence of a $r\approx10$ $\mu$m fiber during a typical experiment.}
\end{figure}

A cylindrical, elastic fiber with a radius of  $\sim10 \ \mu$m is attached to two thin glass capillary pipettes acting as posts. One glass post is mounted on a rotational stepper motor ($1.8^{\circ}$ resolution) and linear actuator, allowing precise control of tension, slack, and degree of twisting in the fiber. The other post, a capillary that acts as a sensitive force transducer, is capable of measuring the tension in the fiber directly. The force transducer pipette is mounted perpendicular to the length of the fiber to facilitate force measurement. A schematic of the experimental setup is shown in Figure~\ref{schem}a).

Both posts are made from glass capillary tubes with a diameter of 1.0~mm (World Precision Instruments, USA). The force transducer is made by pulling a capillary tube with a pipette puller (Narishige, Japan)  to be long ($\sim 1$~cm), thin ($\sim 10 \ \mu$m)  enough  to deflect when tension was applied to the fiber. By calibrating this pipette, its spring constant $k$ ($\sim 0.5$ {N/m}) can be determined, allowing for force measurements as small as hundreds of pN by monitoring the deflection of the pipette $d$ using cross-correlation image analysis~\cite{Colbert2009,Backholm2019MicropipetteFS}. 

The fibers used in this experiment were made from Elastollan (Wacker Chemi AG), a commercially available elastomer with Young's modulus $E = 11 \pm 3 $~MPa which was determined via extensional stress-strain tests performed on several different fibers with $r\approx10 \ \mu$m (not shown). Fibers were made by heating a pellet of Elastollan to $240^{\circ}$~C, dipping a glass pipette into the melt, then rapidly pulling the pipette out of the melt. The resulting fibers have uniform, cylindrical cross-sections with a diameters of $\sim 10 \ \mu$m. The fibers were inspected optically for uniformity, then glued across the posts with a dilute polystyrene-toluene solution. A droplet of the solution was placed at the contact point between the fiber and posts, and as the toluene evaporated, a layer of glassy polystyrene was left which holds the fiber in place. Toluene was selected as the solvent as it selectively dissolves polystyrene and not Elastollan.

In a typical experiment the fiber is rotated at one end by an angle of $\theta_{\circ} = 2 \pi n$ corresponding to $n$ full revolutions [Fig.~\ref{schem}a)], while the fiber of length $L$, is held at an initial tension such that the fiber remains straight and unbuckled. The tension is then released by moving the left post by a distance $D$ and bringing the posts together at a speed of $30\ \mu$m/s [Fig.~\ref{schem}b)]. Strictly, the slack introduced into the fiber $\delta = D+d$, but since the distance the post is moved is $\sim 10^3$ times greater than the deflection of the force sensing pipette, we can take $\delta \approx D$. As the slack is increased, the fiber is observed and found to hockle and writhe as the elastic energy stored in the twisted fiber is converted into bending energy [Fig.~\ref{schem}c)]. An optical image sequence of the fiber during a typical experiment is shown in Fig.~\ref{schem}d). Since the fiber is radially symmetric, twisting in either direction is equivalent and experiments are repeated with an initial twist of $-\theta_{\circ}$ (note that $\theta_{\circ}$ is defined as positive). Repeating the experiment for $\theta_{\circ}$ and $-\theta_{\circ}$ compensates for any effects related to errors in defining $\theta_{\circ}=0$ or radial non-uniformities. The deformations in the fiber as well as the deflection of the force transducer are simultaneously measured with optical microscopy.
\section{\label{sec:Results}Results and Discussion}
\subsection{\label{sec:slack}Formation and removal of a hockle}
We consider the fiber as a slender rod with a large length to width ratio. We follow the argument outlined by Ross, which uses the results of Timoshenko and an analysis of the relevant energies in the system, to derive criteria for hockling related to the tension and torsion within a twisted fiber~\cite{ross_cable_1977,timoshenko}. 

A twisted fiber will hockle and form a loop when the torsional energy stored in the fiber is large enough to overcome any stabilizing tension in the fiber. The bending energy within the resulting loop must be balanced by the work released as as the ends of the fiber are brought together and the fiber untwists. To form a single loop, there are three energy contributions to consider: i) energy is required to bend the fiber into a loop;   $\Delta U_{\text{b}}$, ii) work is released as the two ends holding the fiber are brought closer,   $  \Delta W_{\text{T}}$; and iii) energy stored in the twisted fiber is released, $ \Delta W_{\text{M}}$, because upon formation of a loop the fiber unwinds by one full rotation. 

For the formation of a single loop, the bending energy is calculated assuming the fiber undergoes a linear elastic deformation into a perfect circle with radius $R$,
\begin{equation}\label{eq.Bend1}
    \Delta U_{\text{b}} = \frac{EI}{2R^{2}}2\pi R = \frac{\pi EI}{R},
\end{equation}
where $E$ is Young's modulus and  $I$ is the second area moment of the cylindrical fiber, $I=\pi r^4/4$.
The work done by bringing the two posts together is given by
\begin{align}\label{eq:comp}
    \Delta W_{\text{T}} = -T\Delta D,
\end{align}
where $T$ is tension in the fiber, and $\Delta D$ is the change in the distance between the posts needed to form a loop ($\Delta D$ is defined as positive when the pipettes are brought together and negative as they are pulled apart). Lastly, the work done via untwisting, $\Delta W_{\text{M}}$, is given by,
\begin{align}\label{eq:tor}
    \Delta W_{\text{M}} = 2\pi M,
\end{align}
where $M$ is the twisting moment of the fiber, and $2\pi$ is the angle through which the fiber must unwind to form a single loop. Within the linear elastic regime, $M$ varies linearly with the twist angle and is,
\begin{equation}\label{torque}
    M = \frac{JG\theta_{\circ}}{L},
\end{equation}
where $J$ is the torsional constant for a cylindrical fiber, $J=\pi r^4/2=2I$, and $G$ is the shear modulus of the material (note that this assumption remains valid for large $\theta_{\circ}$ provided $L$ is also large).
The balance between the three energy contributions is then given by, 
\begin{align} \label{eq:fullcrit}
    \frac{\pi EI}{R} + T\Delta D = 2\pi M.
\end{align}
Equation \ref{eq:fullcrit} can be applied to the formation of a loop as well as the removal of a loop, and each case will be considered in turn.

\begin{figure}[]
\centering
\includegraphics[width=0.5\textwidth]{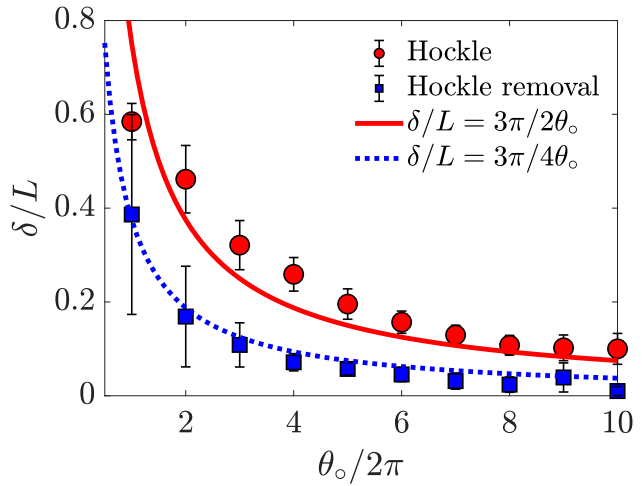}
\caption{Slack $\delta$ normalized by natural length $L$ of fibers with initial twist $\theta_{\circ}$ at the point of hockling (circles, solid line) and removal of the loop (squares, dashed line) and their corresponding theoretical predictions. The data is the average of 10 fibers with lengths varying from $L=$ 6~mm to 300~mm and radii varying from  $r=10 \ \mu$m to $1000 \ \mu$m. Error bars are calculated as the standard deviation of the data. }
\label{fig:hockle_snap}
\end{figure}

In the experiment presented here, a twisted fiber is stabilized against buckling by beginning in a state of tension. As the ends of the fiber are brought together, $T$ decreases rapidly. For small values of $\theta_{\circ}$, the tension $T$ at the point of hockling is minimal and we make the approximation that $T=0$. This allows for Equation \ref{eq:fullcrit} to be simplified:
\begin{equation}\label{hocklecrit}
    M = \frac{EI}{2R}.
\end{equation}
From Equation \ref{torque} and Equation \ref{hocklecrit} we obtain,
\begin{equation}\label{hocrit2}
    \frac{JG\theta_{\circ}}{L} = \frac{EI}{2R}.
\end{equation}
Making the generous assumption that the fiber outside the loop remains straight and all slack in the fiber goes into forming a perfect circle, we can define the slack as ${\delta =2\pi R}$. We note that the assumption of a circular loop results in a small systematic error for small $\theta_{\circ}$ which will be discussed below. Equation \ref{hocrit2} can then be written as,
\begin{equation}
    \delta = \frac{\pi E I}{JG}\frac{L}{\theta_{\circ}}.
\end{equation}
For a cylindrical fiber made from a material with a Poisson ratio of $\nu \approx 0.5$ (typical of elastomers), $E = 2G(1+\nu)$ and $J=2I$, the amount of slack required to form a hockle is given by
\begin{equation}\label{eq:hockle}
    \frac{\delta}{L} = \frac{3\pi}{2\theta_{\circ}}.
\end{equation}
We see from this expression that the amount of slack that needs to be provided in the fiber for loop formation is independent of the material properties of the fiber, and only dependent on geometry and how much the fiber is twisted. This result is to be expected since the formation of a hockle depends on equating the energy to form a loop with the energy stored in the twisted fiber, both of which depend on the modulus. In order to validate this expression, the slack required for a hockle to form for different values of $\theta_{\circ}$ was measured for 10 fibers with lengths varying from $L=$ 6~mm to 300~mm and $r=10 \ \mu$m to $1000 \ \mu$m. The results are plotted in Figure~\ref{fig:hockle_snap}  (circles). A systematic increase in $\delta/L$ for small values of $\theta_{\circ}$ can be seen, which is due to the assumption of a perfectly circular loop and straight fiber outside the loop. The assumption becomes increasingly valid at higher $\theta_{\circ}$. 

Having examined the formation of a hockle, we now turn to the removal of a hockle as the two ends of the fiber are pulled apart. If the ends of  the fiber are pulled apart, $R$ decreases, and the bending energy $\Delta U_{\mathrm{b}}$ increases, until it becomes more energetically favorable to remove the loop and re-twist the fiber. In this case, $T$ is no longer negligible and the work done in pulling apart the ends corresponds to the increase in bending energy in the increasingly small loop. The energy balance in Equation~\ref{eq:fullcrit} then becomes $M=EI/R$. When compared to Equation~\ref{hocklecrit}, there is an extra factor of 2, which results in the prediction of $\delta$ (and size of the loop) when a hockle is removed,
\begin{equation} \label{eq:popout}
    \frac{\delta}{L} = \frac{3\pi}{ \theta_{\circ}}.
\end{equation}

Again, there are no material parameters in the criterion for the removal of a hockle, and the data are shown in Figure~\ref{fig:hockle_snap} (squares). We note that, with the approximation of a circular hockle,  the formation of a hockle requires twice as much slack in the fiber as does the removal of a hockle (compare Equations~\ref{eq:hockle} and \ref{eq:popout}). In other words, the circumference of the loop which forms is twice as large as the circumference of the loop when the loop  is removed. Since neither criteria depend on the material properties of the fiber, Equations \ref{eq:hockle} and \ref{eq:popout} are valid for all uniform elastic rods with circular cross-sections within the linear elastic regime. Effects like sagging due to gravity which would affect large scale systems would modify this model.

\subsection{Plectoneme Growth and Removal}\label{writhe}

In the previous section we investigated the formation and removal of hockles. After a hockle forms in a highly twisted fiber, bringing the ends even closer together can allow a double-helix structure --  a ``plectoneme'' -- to form through a process called writhing. A plectoneme is shown schematically in  Figure~\ref{fig:plectoneme} (see also video in the Supplemental Information). Similar to destabilizing a hockle via tension, a plectoneme can also be destabilized, and this has been done in a number of studies on  DNA using optical and magnetic tweezers \cite{lipfert_magnetic_2010,lipfert2020,Neuman2019,lipfert2018,Li2020,Gerland2020}. 
 In this section we will investigate the growth and removal of plectonemes.
 \begin{figure}
\centering
\includegraphics[width=0.4\textwidth]{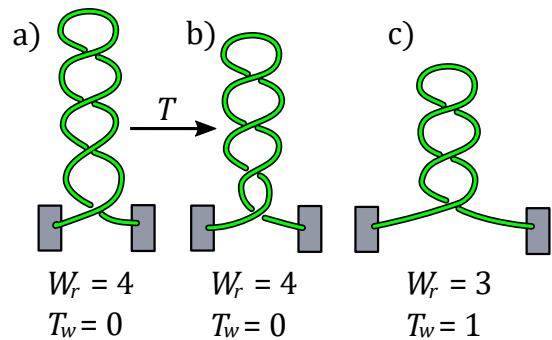}
\caption{Schematic of a plectoneme formed from a fiber with an initial twist angle corresponding to four full rotations ($\theta_\circ=8\pi$). a) The stable plectoneme has minimal tension, and all of the energy stored in the twisted fiber is stored in the bends of the plectoneme. $L_{k} =4$, $T_{w}=0$, and $W_{r}=4$. b) The tension is increased by separating the boundaries, and the bending energy in the loop at the base of the plectoneme increases. c) When the bending energy stored in the base loop is enough to destabilize the base loop, the twist in the fiber increases by one full rotation, and decrease the writhe number by 1:  $T_{w}=1$, and $W_{r}=3$.}
\label{fig:plectoneme}
\end{figure}
 
A plectoneme forms by exchanging the twist in the fiber for loops, resulting in points of self-contact as the fiber winds around itself. The ``linking number'', $L_{k}$, is defined as the sum of the ``twist number'' $T_{w}$ and the ``writhe number'' $W_{r}$, both of which are integers \cite{Junier2020,Riley1981}. $T_{w}$ counts the number of complete $2\pi$ radians of twist in the fiber, while  $W_{r}$ counts the number of self-contacts of the plectoneme~\cite{fraser_theory_2008,kulic_equation_2007,neukirch_analytical_2011,marko_statistical_1995,champneys_multiplicity_1996}. If both ends of the fiber are unable to rotate, then $L_{k} = T_{w} + W_{r}$ is a constant. For example, a fiber with $\theta_{\circ} = 8\pi$ of twist and no self-contacts has $L_{k} = T_{w} = 4$, and $W_{r}=0$. As expected, we observed that as the ends of the fiber were brought together, $T_{w}$ decreased in steps of 1 with simultaneous increases in $W_{r}$. When $W_{r}=1$, we observe a hockle in the fiber and when $W_{r} > 1$, a plectoneme is observed. 

\begin{figure}
\centering
\includegraphics[width=0.5\textwidth]{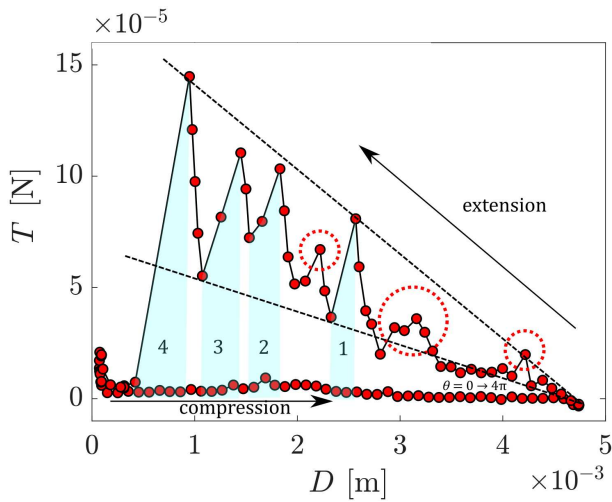}
\caption{$T$ is measured in a twisted fiber with $\theta_{\circ} = 12 \pi$, $L = 6$ mm and $r =$ 18 $\mu$m as its ends are brought together (compression) and then reversed (extension).  As the fiber is compressed, small but distinct tension peaks are observed corresponding to an increase in the writhe number, $W_{r}$, until the tension vanishes within the resolution of the experiment. During extension $T$ and $M$ are initially small, allowing the plectoneme to unwind smoothly. After $W_{r}$ decreases by 2, peaks and valleys in the tension corresponding to a reduction in $W_{r}$ were observed (shaded area, numbered) and were determined by noting the image frames where the plectoneme begins rotating and where it stops rotating. Additional peaks (in dashed circles) are the result of stick-slip events as the fiber moved past itself that are not associated with a change in $W_r$.}
\label{fig:forceschem}
\end{figure}

The results of a typical experiment are shown in Figure~\ref{fig:forceschem} where we plot $T$ as a function of $D$ (see movie at https://youtu.be/iQzRZvDTtDk).  The experiment proceeds as follows: a fiber is initially held under a small tension and twisted by $\theta_{\circ}$. At this point, $D=0$, $W_{r} =0$, and $L_{k} = T_{w} = \theta_{\circ}/2\pi$. One post is then translated with the motorized translation stage, increasing $D$ (this sequence is labelled as \emph{compression} in the figure). Small variations in the tension are observed as a plectoneme forms and  $W_{r}$ increases by forming self-contacts in a quantized manner. The process is then reversed (labelled as \emph{extension} in the figure). Remarkably, a rich tension response emerges with much larger tension required to unwind a plectoneme compared to the formation. We observe large peaks in $T$, followed by sudden drops that are  concurrent with  decreases in the writhe number:  $W_{r} \rightarrow W_{r}-1$. The peaks increase in magnitude as $W_{r}$ decreases, with the highest peak corresponding to the final removal of the hockle.  

To understand the origin of the rich non-monotonic changes in tension, we now investigate what happens when $W_{r} \rightarrow W_{r}-1$.  
Studies have found that a completely frictionless plectoneme experiences increased bending throughout the entire plectoneme structure \cite{purohit_plectoneme_2008,clauvelin_elasticity_2009,Neuman2019}. Other studies, however, find that friction at fiber contact points plays an important role when a plectoneme is pulled apart \cite{min_discontinuous_2018,PhysRevLett.115.118302}. Our model is based on the importance of friction between fiber contacts for the discontinuous plectoneme unwinding. 
In the experiments shown, we observe that typically the self-contact at the base of the plectoneme slips, while the next self-contact sticks. Thus, as the tension is increased, the increase in bending is localized at the base of the plectoneme, and similar to pulling apart a hockle, the rotation of the plectoneme is a sudden event [see Fig.~\ref{fig:plectoneme}b) and c)]. While previous studies have sought to describe the bending energy contained in the plectoneme~\cite{mahasol2005,tanaka_elastic_1985,thompson_helix_1996,champneys_multiplicity_1996,van_der_heijden_instability_2003,starostin_three-dimensional_1996,fraser_theory_2008,gazzola_forward_2018,goss_experiments_2005}, we seek to understand this largely unreported phenomenon of discontinuous plectoneme unwinding: from the experiments we observe that as tension is applied the plectoneme does not continuously unravel, rather, there are sudden and non-monotonic changes in the tension corresponding  to the quantized decrease in  $W_{r}$~\cite{daniels_discontinuities_2009,min_discontinuous_2018,purohit_plectoneme_2008}. 

\begin{figure}
\centering
\includegraphics[width=0.3\textwidth]{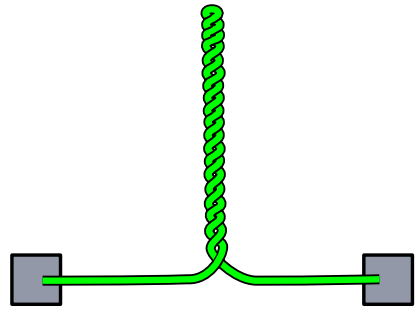}
\caption{Schematic of a plectoneme formed from a fiber with a high twist number and a correspondingly high twisting moment of the fiber $M$}
\label{fig:continuous_plectoneme}
\end{figure}

The schematic shown in Fig.~\ref{fig:plectoneme} illustrates loops in the plectoneme. The size of these loops depend on the twist number: if the energy stored in twists is high compared to the energy required to bend the fiber, then tight loops form; conversely, a low twist number results in a low twisting moment, $M$, and open loops. In this study,  $M$ is low, and the experiments are carried out in a regime where loops form along the plectoneme. We find that the non-monotonic changes in the tension are coincident with the removal of a base loop. It is instructive to consider the limiting case of a fiber dominated by a high value of $M$ with a plectoneme that is tightly wound as shown schematically in Figure~\ref{fig:continuous_plectoneme}, where the dominance of $M$ means that we can ignore the contribution of bending energy.
In this case, if we imagine pulling the boundaries apart then as the tension $T$ increases, so does the twist angle, $\theta$, in the fiber. In fact, any unit of length increase in the boundaries is directly proportional to an increase in $\theta$. Thus $\theta \propto - \Delta D$ for separation of the boundaries (note the negative sign is the result of defining compression as positive). Next, the work done by tension goes into undoing the plectoneme and increasing the twisting energy of the fiber which scales as $\theta^2$. We can then approximate $-T\Delta D \propto \theta^2$, which results in a linear change in the tension with separation of the boundaries like a Hookean spring, $T\propto -\Delta D$, for this continuum approach. Indeed the data shown in Figure~\ref{fig:forceschem} is bound by a linear envelope with the upper and lower boundaries in the tension corresponding to the transition from $W_r$ to $W_r-1$.

During the  $W_r \rightarrow W_r-1$ transition, the tension decreases suddenly from the initial value at the upper bound, to the final value after the transition at the lower-bound, $T_{i}$ and $T_{f}$.  We can understand this from the change in energy immediately before and immediately after the destabilization of the base-loop [see Fig.~\ref{fig:plectoneme}b) and c)]. Prior to destabilization, the tension increases and the work, $ \int T(D) \mathrm{d}D $, increases the energy stored in bending at the base loop. A fraction of that bending energy is released when a loop is removed and a twist is added to the fiber: $W_r \rightarrow W_r-1$ and $T_w \rightarrow T_w+1$, resulting in a sudden decrease in the tension from $T_i$ to $T_f$. As the tension decreases, the distance between the boundaries increases by some length roughly equal to the slack created through the loss of the base-loop, $l$. Thus we have $(T_f-T_i) l \sim U_{b,f}- U_{b,i}$, where $U_{b,i}$ and $U_{b,f}$ are the bending energies before and after the destabilization of the base-loop. However, since the loss in bending energy is transferred into twist energy, we can write
 \begin{equation}
   - (T_f-T_i) l \sim \frac{JG}{L}(\theta_f^2-\theta_i^2)=\frac{JG}{L}[4\pi( \theta_i+ \pi)].
   \label{e:tension}
 \end{equation}
 We see that finally we obtain a linear dependence on the twist angle which bounds the maxima and minima in the tension given by this expression. Furthermore we see from Eq.~\ref{e:tension}, that $T_i-T_f$ increases with the degree of twist in the fiber, which is validated by the data since $\theta$ increases in Figure~\ref{fig:forceschem} as $D$ decreases. Making the assumption, as in the continuum model above, that the variation in $\theta$ is linear in $D$, since each loop of the plectoneme is the same size, explains why the upper bound and lower bound are also linear in $D$. 
 
 We noted above that the discontinuous change in the tension associated with the quantized nature of destabilizing a plectoneme has been reported by few studies in the literature, while the sudden drop in tension associated with destabilizing a hockle is well known \cite{daniels_discontinuities_2009,min_discontinuous_2018,purohit_plectoneme_2008}. We attribute our success in measuring this effect to the small scale of our experiment. Since the magnitude of the tension peaks are linearly dependent on the twist angle $\theta$ (see Eq.~\ref{e:tension}), experiments for which $\theta_{\circ}$ is small may not exhibit large peaks until the final removal of a hockle. However, to stay within the linear elastic regime, $\theta_{\circ}/L$ must remain small. Because our fibers are exceedingly slender, we are able to perform experiments with relatively large $\theta_{\circ}$ while still remaining in the linear elastic regime. Finally, because our fibers are so small, sagging due to gravity is eliminated, facilitating the study of plectoneme formation and unravelling.  
 \section{Conclusion}\label{conclusion}
 We  have extended the energy analysis of Ross and Yabuta~\cite{ross_cable_1977,yabuta_submarine_1984} to predict the point at which the hockle is formed and destabilized, and validated both criteria with precise micron-scale experiments. The idealized system was also used to explore the formation and removal of plectonemes, observing multiple instabilities associated with the change in number of self-contacts within the plectoneme. The changes in tension observed with the experiments are well described by a simple model.

\bibliographystyle{unsrt}
\bibliography{hockwrithe}

\begin{thebibliography}{10}

\bibitem{yu_bifurcations_2018}
T.~Yu and J.A. Hanna.
\newblock Bifurcations of buckled, clamped anisotropic rods and thin bands
  under lateral end translations.
\newblock {\em Journal of the Mechanics and Physics of Solids}, February 2018.

\bibitem{champneys_multiplicity_1996}
A.~R. Champneys and J.~M.~T. Thompson.
\newblock A multiplicity of localized buckling modes for twisted rod equations.
\newblock {\em Proceedings of the Royal Society of London A},
  452(1954):2467--2491, 1996.

\bibitem{thompson_helix_1996}
J.~M.~T. Thompson and A.~R. Champneys.
\newblock From helix to localized writhing in the torsional post-buckling of
  elastic rods.
\newblock {\em Proceedings of the Royal Society of London A},
  452(1944):117--138, 1996.

\bibitem{coyne_analysis_1990}
J.~Coyne.
\newblock Analysis of the formation and elimination of loops in twisted cable.
\newblock {\em IEEE Journal of Oceanic Engineering}, 15(2):72--83, 1990.

\bibitem{dias_wunderlich_2015}
M.~A. Dias and B.~Audoly.
\newblock “{Wunderlich}, {Meet} {Kirchhoff}”: {A} {General} and {Unified}
  {Description} of {Elastic} {Ribbons} and {Thin} {Rods}.
\newblock {\em Journal of Elasticity}, 119(1-2):49--66, April 2015.

\bibitem{van_der_heijden_instability_2003}
G.H.M. van~der Heijden, S.~Neukirch, V.G.A. Goss, and J.M.T. Thompson.
\newblock Instability and self-contact phenomena in the writhing of clamped
  rods.
\newblock {\em International Journal of Mechanical Sciences}, 45(1):161--196,
  January 2003.

\bibitem{van_der_heijden_lock-tape-like_1999}
G.~H.~M. Van~der Heijden and J.~M.~T. Thompson.
\newblock Lock-on to tape-like behaviour in the torsional buckling of
  anisotropic rods.
\newblock In {\em Localization {And} {Solitary} {Waves} {In} {Solid}
  {Mechanics}}, pages 133--156. World Scientific, 1999.

\bibitem{clauvelin_matched_2009}
N.~Clauvelin, B.~Audoly, and S.~Neukirch.
\newblock Matched asymptotic expansions for twisted elastic knots: a
  self-contact problem with non-trivial contact topology.
\newblock {\em Journal of the Mechanics and Physics of Solids},
  57(9):1623--1656, 2009.

\bibitem{gazzola_forward_2018}
M.~Gazzola, L.~H. Dudte, A.~G. McCormick, and L.~Mahadevan.
\newblock Forward and inverse problems in the mechanics of soft filaments.
\newblock {\em Royal Society Open Science}, 5(6):171628, June 2018.

\bibitem{goss_experiments_2005}
V.~G.~A. Goss, G.~H.~M. van~der Heijden, J.~M.~T. Thompson, and S.~Neukirch.
\newblock Experiments on snap buckling, hysteresis and loop formation in
  twisted rods.
\newblock {\em Experimental Mechanics}, 45(2):101--111, April 2005.

\bibitem{stump_hockling_2000}
D.~M. Stump.
\newblock The hockling of cables: a problem in shearable and extensible rods.
\newblock {\em International journal of solids and structures}, 37(3):515--533,
  2000.

\bibitem{ross_cable_1977}
A.~L. Ross.
\newblock Cable kinking analysis and prevention.
\newblock {\em Journal of Engineering for Industry}, 99(1):112--115, 1977.

\bibitem{mahasol2005}
A.~Ghatak and L.~Mahadevan.
\newblock Solenoids and plectonemes in stretched and twisted elastomeric
  filaments.
\newblock {\em Physical Review Letters}, 95:057801, Jul 2005.

\bibitem{nizette_towards_1999}
M.~Nizette and A.~Goriely.
\newblock Towards a classification of {Euler}–{Kirchhoff} filaments.
\newblock {\em Journal of Mathematical Physics}, 40(6):2830--2866, June 1999.

\bibitem{neukirch_classification_2002}
S.~Neukirch and M.~E. Henderson.
\newblock Classification of the spatial equilibria of the clamped elastica:
  {Symmetries} and zoology of solutions.
\newblock {\em Journal of Elasticity}, 68(1-3):95--121, 2002.

\bibitem{PhysRevLett.115.118302}
M.~K. Jawed, P.~Dieleman, B.~Audoly, and P.~M. Reis.
\newblock Untangling the mechanics and topology in the frictional response of
  long overhand elastic knots.
\newblock {\em Phys. Rev. Lett.}, 115:118302, Sep 2015.

\bibitem{wada_structural_2016}
H.~Wada.
\newblock Structural mechanics and helical geometry of thin elastic composites.
\newblock {\em Soft Matter}, 12(35):7386--7397, 2016.

\bibitem{fraser_theory_2008}
W.~B. Fraser and G.~H.~M. van~der Heijden.
\newblock On the theory of localised snarling instabilities in false-twist yarn
  processes.
\newblock {\em Journal of Engineering Mathematics}, 61(1):81--95, May 2008.

\bibitem{ermolaeva_hockling_2008}
N.~S. Ermolaeva, J.~Regelink, and M.~P.M. Krutzen.
\newblock Hockling behaviour of single- and multiple-rope systems.
\newblock {\em Engineering Failure Analysis}, 15(1-2):142--153, January 2008.

\bibitem{habibi_coiling_2007}
M.~Habibi, N.~M. Ribe, and D.~Bonn.
\newblock Coiling of {Elastic} {Ropes}.
\newblock {\em Physical Review Letters}, 99(15), October 2007.

\bibitem{yabuta_submarine_1984}
T.~Yabuta.
\newblock Submarine {Cable} {Kink} {Analysis}.
\newblock {\em Bulletin of JSME}, 27(231):1821--1828, 1984.

\bibitem{kojima_cable_1982}
N.~Kojima.
\newblock Cable {Kink} {Analysis}; {Cable} {Loop} {Stability} {Under}
  {Tension}.
\newblock {\em Journal of Applied Mechanics}, 49:585, 1982.

\bibitem{marko_competition_2012}
J.~F. Marko and S.~Neukirch.
\newblock Competition between curls and plectonemes near the buckling
  transition of stretched supercoiled {DNA}.
\newblock {\em Physical Review E}, 85(1), January 2012.

\bibitem{fraser_equilibrium_1998}
W.~B. Fraser and D.~M. Stump.
\newblock The equilibrium of the convergence point in two-strand yarn plying.
\newblock {\em International journal of solids and structures},
  35(3-4):285--298, 1998.

\bibitem{purohit_plectoneme_2008}
P.~K. Purohit.
\newblock Plectoneme formation in twisted fluctuating rods.
\newblock {\em Journal of the Mechanics and Physics of Solids},
  56(5):1715--1729, May 2008.

\bibitem{clauvelin_elasticity_2009}
N.~Clauvelin, B.~Audoly, and S.~Neukirch.
\newblock Elasticity and {Electrostatics} of {Plectonemic} {DNA}.
\newblock {\em Biophysical Journal}, 96(9):3716--3723, May 2009.

\bibitem{neukirch_analytical_2011}
S.~Neukirch and J.~F. Marko.
\newblock Analytical {Description} of {Extension}, {Torque}, and {Supercoiling}
  {Radius} of a {Stretched} {Twisted} {DNA}.
\newblock {\em Physical Review Letters}, 106(13), April 2011.

\bibitem{kulic_equation_2007}
I.~M. Kuli\'{c}, H.~Mohrbach, R.~Thaokar, and H.~Schiessel.
\newblock Equation of state of looped {DNA}.
\newblock {\em Physical Review E}, 75(1), January 2007.

\bibitem{marko_global_2013}
J.~F. Marko and S.~Neukirch.
\newblock Global force-torque phase diagram for the {DNA} double helix:
  {Structural} transitions, triple points, and collapsed plectonemes.
\newblock {\em Physical Review E}, 88(6), December 2013.

\bibitem{smith_predicting_2008}
M.L. Smith and T.J. Healey.
\newblock Predicting the onset of {DNA} supercoiling using a non-linear
  hemitropic elastic rod.
\newblock {\em International Journal of Non-Linear Mechanics},
  43(10):1020--1028, December 2008.

\bibitem{daniels_discontinuities_2009}
B.~C. Daniels, S.~Forth, M.~Y. Sheinin, M.~D. Wang, and J.~P. Sethna.
\newblock Discontinuities at the {DNA} supercoiling transition.
\newblock {\em Physical Review E}, 80(4), October 2009.

\bibitem{chamekh_stability_2014}
M.~Chamekh, S.~Mani-Aouadi, and M.~Moakher.
\newblock Stability of elastic rods with self-contact.
\newblock {\em Computer Methods in Applied Mechanics and Engineering},
  279:227--246, September 2014.

\bibitem{brutzer_energetics_2010}
H.~Brutzer, N.~Luzzietti, D.~Klaue, and R.~Seidel.
\newblock Energetics at the {DNA} {Supercoiling} {Transition}.
\newblock {\em Biophysical Journal}, 98(7):1267--1276, April 2010.

\bibitem{silverberg_3d_2012}
J.~L. Silverberg, R.~D. Noar, M.~S. Packer, M.~J. Harrison, C.~L. Henley,
  I.~Cohen, and S.~J. Gerbode.
\newblock {3D imaging and mechanical modeling of helical buckling in {Medicago}
  truncatula plant roots}.
\newblock {\em Proceedings of the National Academy of Sciences},
  109(42):16794--16799, 2012.

\bibitem{gerbode_how_2012}
S.~J. Gerbode, J.~R. Puzey, A.~G. McCormick, and L.~Mahadevan.
\newblock How the {Cucumber} {Tendril} {Coils} and {Overwinds}.
\newblock {\em Science}, 337(6098):1087--1091, August 2012.

\bibitem{tanaka_elastic_1985}
F.~Tanaka and H.~Takahashi.
\newblock Elastic theory of supercoiled {DNA}.
\newblock {\em The Journal of Chemical Physics}, 83(11):6017--6026, December
  1985.

\bibitem{coleman_theory_1995}
B.~D. Coleman, I.~Tobias, and D.~Swigon.
\newblock Theory of the influence of end conditions on self‐contact in {DNA}
  loops.
\newblock {\em The Journal of Chemical Physics}, 103(20):9101--9109, November
  1995.

\bibitem{hoffman_link_2003}
K.~A. Hoffman, R.~S. Manning, and J.~H. Maddocks.
\newblock Link, twist, energy, and the stability of {DNA} minicircles.
\newblock {\em Biopolymers: Original Research on Biomolecules}, 70(2):145--157,
  2003.

\bibitem{starostin_three-dimensional_1996}
E.~L. Starostin.
\newblock Three-dimensional shapes of looped {DNA}.
\newblock {\em Meccanica}, 31(3):235--271, 1996.

\bibitem{gromiha_anisotropic_1996}
M.~M. Gromiha, M.~G. Munteanu, A.~Gabrielian, and S.~Pongor.
\newblock Anisotropic elastic bending models of {DNA}.
\newblock {\em Journal of Biological Physics}, 22(4):227--243, 1996.

\bibitem{cherstvy_looping_2011}
A.~G. Cherstvy.
\newblock Looping charged elastic rods: applications to protein-induced {DNA}
  loop formation.
\newblock {\em European Biophysics Journal}, 40(1):69--80, January 2011.

\bibitem{dobrovolskaia_twist_2010}
I.~V. Dobrovolskaia, M.~Kenward, and G.~Arya.
\newblock Twist {Propagation} in {Dinucleosome} {Arrays}.
\newblock {\em Biophysical Journal}, 99(10):3355--3364, November 2010.

\bibitem{goyal_nonlinear_2005}
S.~Goyal, N.C. Perkins, and C.L. Lee.
\newblock Nonlinear dynamics and loop formation in {Kirchhoff} rods with
  implications to the mechanics of {DNA} and cables.
\newblock {\em Journal of Computational Physics}, 209(1):371--389, October
  2005.

\bibitem{lipfert_magnetic_2010}
J.~Lipfert, J.~W.~J. Kerssemakers, T.~Jager, and N.~H. Dekker.
\newblock Magnetic torque tweezers: measuring torsional stiffness in {DNA} and
  {RecA}-{DNA} filaments.
\newblock {\em Nature Methods}, 7(12):977--980, December 2010.

\bibitem{mosconi_measurement_2009}
F.~Mosconi, J.~F. Allemand, D.~Bensimon, and V.~Croquette.
\newblock Measurement of the {Torque} on a {Single} {Stretched} and {Twisted}
  {DNA} {Using} {Magnetic} {Tweezers}.
\newblock {\em Physical Review Letters}, 102(7), February 2009.

\bibitem{marko_statistical_1995}
J.~F. Marko and E.~D. Siggia.
\newblock Statistical mechanics of supercoiled {DNA}.
\newblock {\em Physical Review E}, 52(3):2912, 1995.

\bibitem{stump_writhing_1998}
D.~M. Stump, W.~B. Fraser, and K.~E. Gates.
\newblock The writhing of circular cross–section rods: undersea cables to
  {DNA} supercoils.
\newblock {\em Proceedings of the Royal Society of London. Series A:
  Mathematical, Physical and Engineering Sciences}, 454(1976):2123--2156, 1998.

\bibitem{ganji_intercalation-based_2016}
M.~Ganji, S.~H. Kim, J.~van~der Torre, E.~Abbondanzieri, and C.~Dekker.
\newblock Intercalation-{Based} {Single}-{Molecule} {Fluorescence} {Assay} {To}
  {Study} {DNA} {Supercoil} {Dynamics}.
\newblock {\em Nano Letters}, 16(7):4699--4707, July 2016.

\bibitem{studart_bioinspired_2014}
A.~R. Studart and R.~M. Erb.
\newblock Bioinspired materials that self-shape through programmed
  microstructures.
\newblock {\em Soft Matter}, 10(9):1284--1294, 2014.

\bibitem{hu_buckling-induced_2015}
N.~Hu and R.~Burgue\~{n}o.
\newblock Buckling-induced smart applications: recent advances and trends.
\newblock {\em Smart Materials and Structures}, 24(6):063001, June 2015.

\bibitem{Colbert2009}
M.~J. Colbert, A.~N. Raegen, C.~Fradin, and K.~Dalnoki-Veress.
\newblock Adhesion and membrane tension of single vesicles and living cells
  using a micropipette-based technique.
\newblock {\em The European Physical Journal E}, 30(2):117, Sep 2009.

\bibitem{Backholm2019MicropipetteFS}
M~Backholm and O~B{\"a}umchen.
\newblock Micropipette force sensors for in vivo force measurements on single
  cells and multicellular microorganisms.
\newblock {\em Nature Protocols}, 14:594--615, 2019.

\bibitem{timoshenko}
S.~P. Timoshenko.
\newblock {\em Theory of elasticity}.
\newblock Engineering societies monographs. McGraw-Hill, 1987.

\bibitem{lipfert2020}
A.~L\"{o}f, P.~U. Walker, S.~M. Sedlak, S.~Gruber, Obser T., M.~A. Brehm,
  M.~Benoit, and J.~Lipfert.
\newblock {Multiplexed protein force spectroscopy reveals equilibrium protein
  folding dynamics and the low-force response of von Willebrand factor}.
\newblock {\em Proceedings of the National Academy of Sciences},
  116(38):18798--18807, 2019.

\bibitem{Neuman2019}
A.~Dittmore, J.~Silver, and K.~C. Neuman.
\newblock {Kinetic Pathway of Torsional DNA Buckling}.
\newblock {\em Journal of Physical Chemistry B.}, 122(49):11561--11570, 2019.

\bibitem{lipfert2018}
P.~U. Walker, W.~Vanderlinden, and J.~Lipfert.
\newblock {Dynamics and energy landscape of DNA plectoneme nucleation}.
\newblock {\em Physical Review E}, 98:042412--042425, 2018.

\bibitem{Li2020}
H.~Wang and H.~Li.
\newblock {Mechanically tightening, untying and retying a protein trefoil knot
  by single-molecule force spectroscopy}.
\newblock {\em Chemical Science}, (11):12512--12521, 2020.

\bibitem{Gerland2020}
K.~Ott, L.~Martini, J.~Lipfert, and U.~Gerland.
\newblock {Dynamics of the Buckling Transition in Double-Stranded DNA and RNA}.
\newblock {\em Biophysical Journal}, 118:1690--1701, 2020.

\bibitem{Junier2020}
M.~Joyeux and I.~Junier.
\newblock {Requirements for DNA-Bridging Proteins to Act as Topological
  Barriers of the Baterial Genome}.
\newblock {\em Biophysical Journal}, 119:1215--1225, 2020.

\bibitem{Riley1981}
A.~Worcel, S.~Strogatz, and D.~Riley.
\newblock {Structure of chromatin and the linking number of DNA}.
\newblock {\em Proceedings of the National Academy of Sciences},
  78(3):1461--1465, 1981.

\bibitem{min_discontinuous_2018}
Y.~Min and P.~K. Purohit.
\newblock Discontinuous growth of {DNA} plectonemes due to atomic scale
  friction.
\newblock {\em Soft Matter}, 14(37):7759--7770, 2018.

\end{thebibliography}

\newpage
\end{document}